\documentclass{camera}
\usepackage{psfig}
\usepackage{graphics}

\begin{document}

\title{ THE DIFFUSE INTERSTELLAR BAND AT 8620 \AA:\\
A GOOD REDDENING TRACER FOR GAIA}

\author{Ulisse Munari}

\organization{\sl Padova and Asiago Astronomical Observatories, \ \ Italy} 

\maketitle

\noindent
{\bf Abstract}. We report about a suprisingly good
correlation between the equivalent width of the diffuse interstellar band
(DIB) at 8620 \AA\ and the interstellar reddening. Such a correlation offers
bright prospects in using the 8620 \AA\ DIB as a tracer of the extinction
through the Galaxy in the context of the GAIA mission by ESA.

\vskip 1.0 cm
\noindent

The GAIA mission planned by ESA is aimed to provide astrometry in the 10~{\sl micro-}arcsec 
regime and multiband photometry for all stars down to mag
20. The mission Red Book prepared by ESA plans GAIA to host a spectrograph to measure radial
velocities and therefore to provide the 6$^{th}$ component of the
phase-space coordinates for all stars brighter than V$\sim$16 (the exact
limit will be set by final optical design, spacecraft scanning law, global
throughput and S/N threshold to trigger the detection during crossing of the
focal plane, dimensions of the latter, on-board data analysis and telemetry
rate, etc.). In the current GAIA design as base-lined in the Red Book,
the spectroscopic observations will be carried out over the 8500-8750 \AA\
region at $\lambda$/$\bigtriangleup\lambda$ =20,000 resolving power
(corresponding to 0.25 \AA/pix and a 2 pix FWHM in the PSF sense).

\begin{figure}
\centerline{\psfig{file=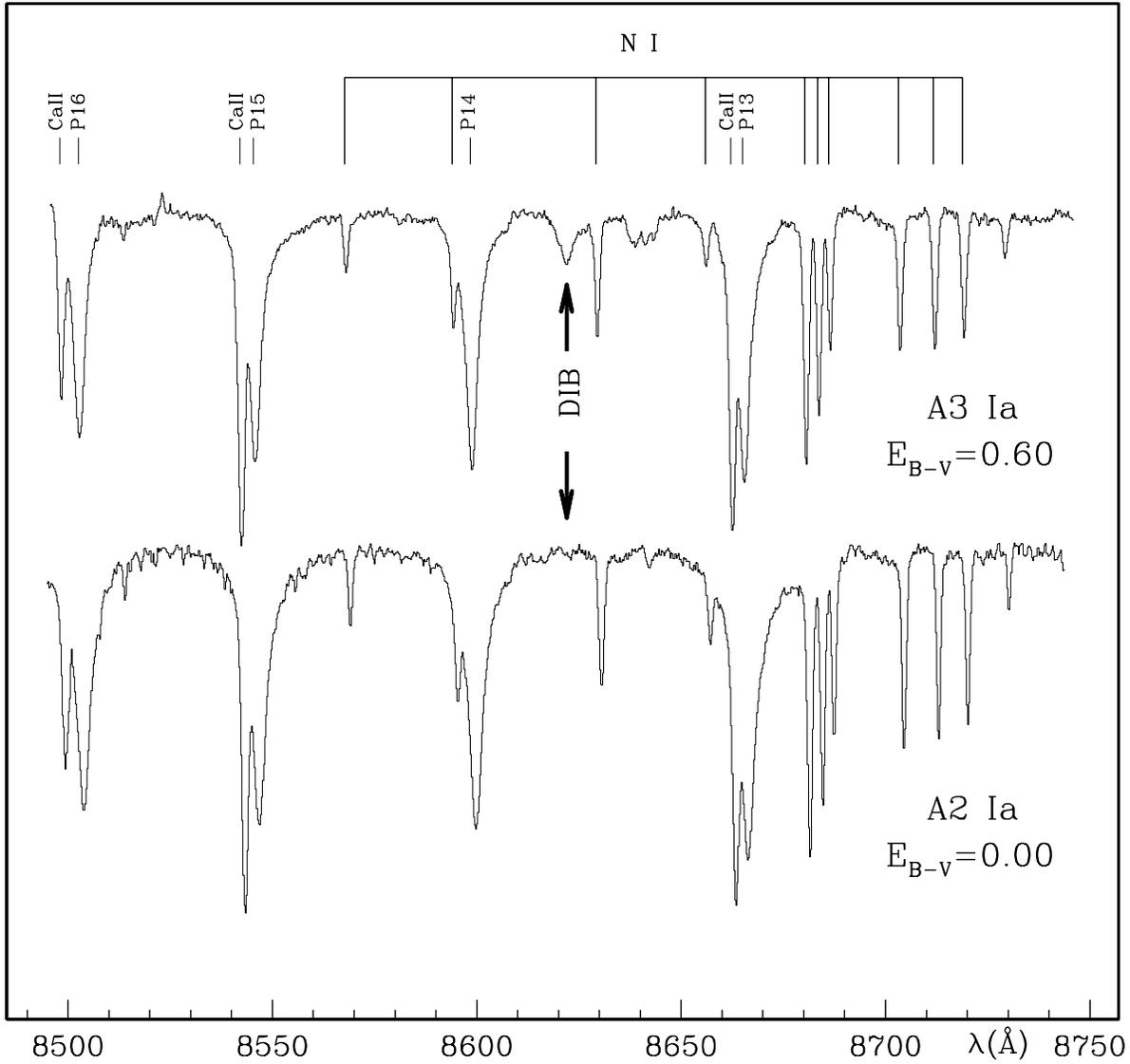,width=16cm}}
\caption[]{\sl Comparison between un-reddened and a moderately reddened
early A-type supergiants (HD 197345 and HD 223385, respectively). 
The strongest stellar absorption lines and the diffuse interstellar band at 8620 \AA\ 
are identified (adapted from Munari \& Tomasella 1999).}
\end{figure}

\begin{figure}
\centerline{\psfig{file=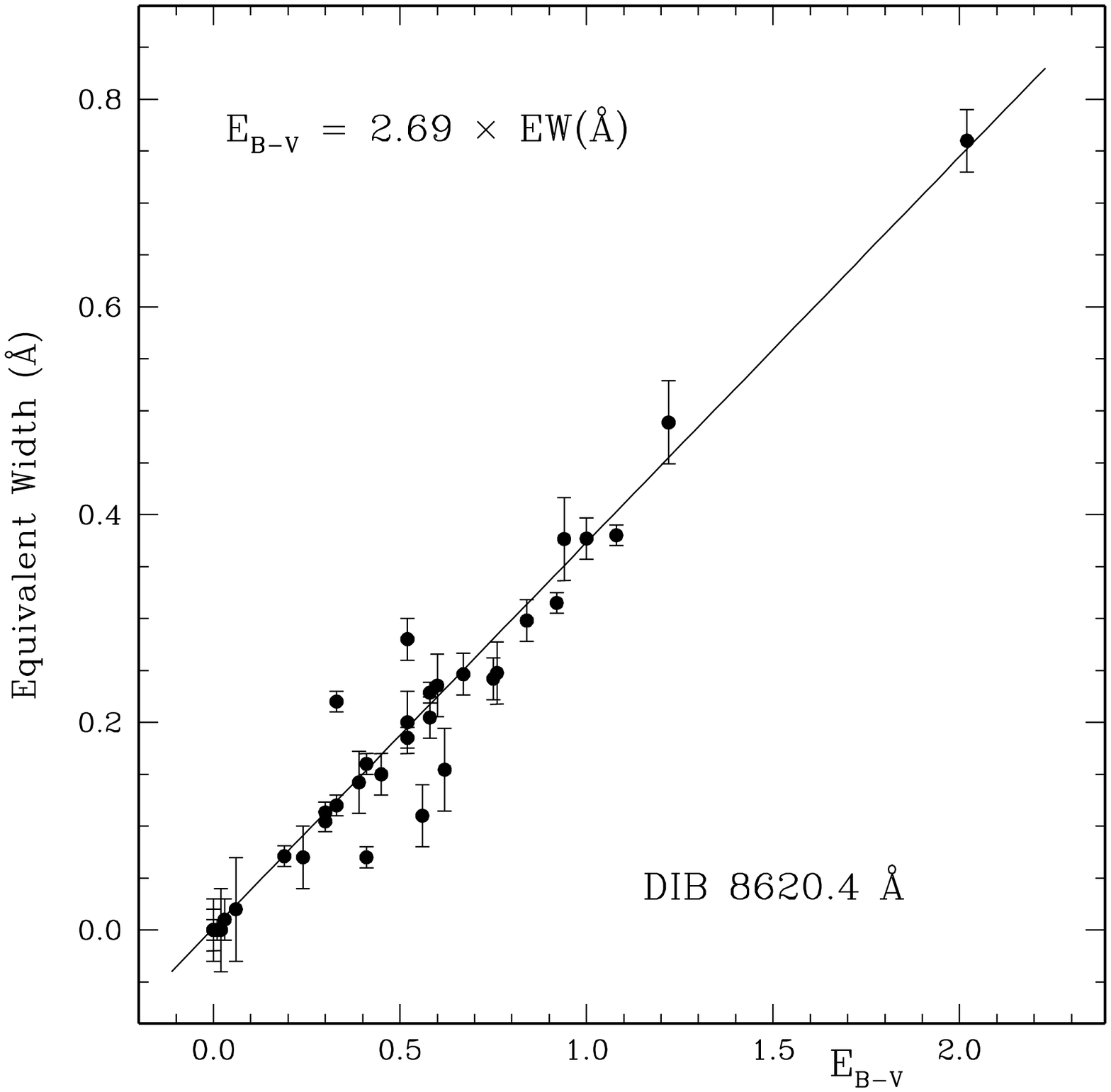,width=16cm}}
\caption[]{\sl Correlation between the equivalent width of the diffuse
interstellar band at 8620 \AA\ and the reddening for stars widely spread 
in galactic coordinates and distance from the Sun (adapted from Munari 1999
and Munari et al. 1999).}
\end{figure}

The astrophysical outlines of the GAIA mission are discussed by Gilmore et
al. (1998) and an outlook of the GAIA payload and spacecraft is presented by
M\'erat et al. (1999). The goals of GAIA spectroscopy and the performance of the
8500-8750 \AA\ region are discussed by Munari (1999). Atlases of real and
synthetic spectra in this region at the resolution base-lined for GAIA may
be found in Munari \& Tomasella (1999) and Munari \& Castelli (2000).

No ion abundant in the interstellar space has resonant lines in the 8500-8750
\AA. There are however some {\sl Diffuse Interstellar Bands} (DIBs). The strongest 
one is at $\lambda$ 8620 \AA. It has been so far very poorly studied in
literature. An example of the band from real spectra secured with the
Asiago Echelle spectrograph is offered in Figure~1.

Given the extreme high accuracy of the astrometry and photometry carried out
by GAIA (distances accurate to 10\% at 10 kpc at V=15 mag, photometry
accurate to 0.001 mag for the majority of the 10$^9$ target stars), a
measure of the DIBs in the 8500-8750 \AA\ by the on-board spectrograph could
provide an accurate 3-D map of the distribution of the DIB carriers along
{\sl any} line of sight in the Galaxy (and the nearby dwarf satellite
galaxies which brightest stars will be reachable by GAIA).

A topic of the highest relevance is a proper handling of the reddening while
reducing and interpreting the GAIA observations. Will the measure of the
DIBs be of any assistance in this matter? It obviously depends
on if a relation between reddening (here written as $E_{B-V}$) 
and some DIB-related quantity (like the equivalent width ($EW$) or the central
depth) exists and how tight is the correlation. DIBs in the optical
range have been generally reported to correlate poorly with reddening. In
There may be really a problem with some DIBs, but in other cases it
is quite possible that systematic observational errors, a naive approach to
data handling or even the use of a wrong extinction law for a given
direction in the Galaxy have artificially resulted in an apparently poor
correlation.

Munari (1999) has shown how a relation exist between $EW$ and
$E_{B-V}$ for the 8620 \AA\ DIB, that was calibrated as 
$E_{B-V} = 2.63 \times $EW($\AA$) on the base of 11 stars observed with the
Asiago Echelle spectrograph. In the meantime Munari et al. (2000) have continued the
observations and expanded the sample to 37 stars widely distributed 
in galactic coordinates and distances. The result from this enlarged sample
is (see Figure~2)

\[ E_{B-V} \ = 2.69\ \times \ EW({\rm \AA})\]

\noindent
which nicely confirms the earlier results. The tight correlation shown
in Figure~2 definitively proves the utility of the 8620
\AA\ DIB as a reddening tracer and meter, even at low reddening values.
However, the exact slope of the relation can be expected to depend upon
the properties of the interstellar material, which change with galactic
coordinates and with distance along a given line of sight. Therefore,
the most general relation would take the form 

\[ E_{B-V} \ = \ \alpha(l,b,{\rm D})\ \times \ EW({\rm \AA})\]

The fine-grid calibration of $\alpha(l,b,{\rm D})$ will be performed by GAIA
itself. In the meantime, we are carrying out additional observations that
will be discussed by Munari et al. (1999; see also Moro \& Zwitter in these
Proceedings), to derive $\alpha(l,b,{\rm D})$ along selected lines of sight in the
Galaxy, to the specific aim to investigate if the relation of Figure~2 (the
mean over many different line of sights) will sharpen further when applied
to narrower and narrower $l,b$ cones, which will make it more and more
useful as a reddening meter.\\

\noindent
{\bf References}
\bigskip

\noindent
Gilmore G., Perryman M., Lindegren L. Favata F., Hoeg E., Lattanzi M., Luri X., 

Mignard F., Roeser S., de Zeeuw P.T., 1998, Proc SPIE Conf. 3350, p. 541\\
Harbig G.H., 1995, ARA\&A 33, 19\\
Merat P., Safa F., Camus J.P., Pace O., Perryman M.A.C. 1999, in Proc. of the 

ESA Leiden Workshop on GAIA, 23-27 Nov 1998, Baltic Astronomy, 8, 1\\
Munari U., 1999, in Proc. of the ESA Leiden Workshop on GAIA, 

23-27 Nov 1998, Baltic Astronomy, 8, 73\\
Munari U., Castelli F., 2000, A\&AS 141, 141\\
Munari U., Tomasella L., 1999, A\&AS 137, 521\\
Munari U., Zwitter T., Tomasella L., Moro D., Porceddu I. 2000, in preparation\\
\end{document}